\documentclass[twocolumn,aps,pra,superscriptaddress,longbibliography]{revtex4-2}
\usepackage{amsmath}
\usepackage{enumitem}
\usepackage{amssymb}
\usepackage{bm}
\usepackage{hyperref}
\usepackage[dvipsnames]{xcolor}
\usepackage[utf8]{inputenc}
\usepackage{graphicx}
\usepackage{dcolumn}
\usepackage{bm}
\usepackage{physics}
\usepackage{wrapfig}
\usepackage{xcolor}
\usepackage[normalem]{ulem}

\begin{document}

\title{Topological finite size effect in one-dimensional chiral symmetric systems}

\date{\today}
\begin{abstract}
Topological phases of matter have been widely studied for their robustness against impurities and disorder. The broad applicability of topological materials relies on the reliable transition from idealized, mathematically perfect models to finite, real-world implementations. In this paper, we explore the effects of finite size and disorders on topological properties. 
We propose a new criterion for characterizing finite topological systems based on the bulk conductivity of topological edge modes. We analyze the behavior of bulk conductivity and real space topological invariants both analytically and numerically for the family of SSH models. We show that our approach offers practical insights for topology determination in contemporary intermediate scale experimental applications.

\end{abstract}
\author{Guliuxin Jin}
\email{g.jin@tudelft.nl}
\affiliation{QuTech and Kavli Institute of Nanoscience, Delft University of Technology, 2628 CJ Delft, the Netherlands}%
\author{D. O. Oriekhov}
\affiliation{QuTech and Kavli Institute of Nanoscience, Delft University of Technology, 2628 CJ Delft, the Netherlands}
\author{Lukas Johannes Splitthoff}
\affiliation{QuTech and Kavli Institute of Nanoscience, Delft University of Technology, 2628 CJ Delft, the Netherlands}%
\author{Eliska Greplova}%
\affiliation{QuTech and Kavli Institute of Nanoscience, Delft University of Technology, 2628 CJ Delft, the Netherlands}%

\maketitle


\section{Introduction}

The discovery of topological phases of matter has revolutionized contemporary condensed matter and solid state physics~\cite{thouless1982quantized, kane2005quantum, bernevig2006quantum, wang2017topological, acin2018quantum, keimer2017physics,wan2011topological, soluyanov2015type, qi2009time,fu2010odd,sasaki2011topological,qi2010chiral}. 
Significant attention has been given to the deep connection between physics and mathematics, particularly in the construction of topological invariants of different phases~\cite{asboth2016short,ryu2010topological,berry1984quantal}. While topological invariants have been originally formulated in the momentum space~\cite{ryu2010topological,berry1984quantal,asboth2016short,hasan2010colloquium}, there has been much interest in real space representations of these invariants, such that the topology can be readily identified and leveraged in finite-sized and disordered system setting~\cite{mondragon2014topological,song2014aiii,prodan2016bulk,lin2021real,meier2018observation,zhang2022generalized}.Recent experimental progress in realizing topological modes highlights the need for easily applicable criteria to determine whether the observed model retains the desired topological properties or if these are compromised by system size or noise.~\cite{Kanungo_2022,splitthoff2024gate,de2019observation,kiczynski2022engineering,jouanny2024band,culcer2020transport,kim2021quantum,mei2018robust,zheng2022engineering,bandres2018topological}. 

The robust topological protection of edge states makes topological materials essential for quantum information and quantum transport applications~\cite{jin2023topological,kim2021quantum,freedman2003topological,dennis2002topological,stern2013topological,ten2024two}. It is crucial to analyze the model's topology under finite size and to keep track of edge states contributions. In one-dimensional chiral models, the edge states are localized at different edges  in large systems, while the bulk remains insulating. However, in smaller systems, edge states can hybridize due to the significant overlap between them through the bulk, resulting in the formation of new symmetric and antisymmetric edge states. This phenomenon is referred to as the finite size effect~\cite{chen2019finite,ozawa2014topological,varney2011topological,jiang2020topological,kalozoumis2018finite}. In large systems, the effect is suppressed by exponentially small overlap of states~\cite{asboth2016short}.

In order to assign the topological invariants to finite systems properly, the real space representation of winding number (RSWN) was introduced \cite{mondragon2014topological,rakovszky2017detecting,prodan2016bulk,lin2021real,zhang2022generalized}. This formulation arises from substituting momentum derivatives in the Berry connection with a commutator involving the coordinate operator in real space. However, it is important to note that the existing RSWN expressions do not fully account for the potential overlap between edge states that may occur due to the finite size of the system. As a result, there are instances where this formula yields anomalous winding number values. One possible way to resolve this issue was suggested in Ref.~\cite{song2019non} involving the truncation of edge sites, effectively mitigating the contributions from edge states.

In this work, we present a new approach to identify the real space topology. We investigate the contribution of finite size effect to real-space topological invariants and topological phase transition. Our analysis reveals that relying solely on the real space topological invariant is insufficient, as it may lead to misjudgments caused by finite size effects. In order to solve this issue, we develop criteria to characterize the real space topological phase transition through the bulk conductivity of the mid-gap edge states. Specifically we utilize real-space invariant in conjunction with the bulk conductivity of topological mid-gap edge states to characterize the topology. We demonstrate our criteria in two types of one dimensional chiral symmetric systems: the Su-Schriefer-Heeger (SSH) model~\cite{su1979solitons,hasan2010colloquium,asboth2016short} and the extended SSH model~\cite{perez2018ssh}. Our criteria provide a robust framework for determining the topology relevant to real-world experimental applications, such as bosonic SSH model realized via Rydberg atoms~\cite{de2019observation}, the in-situ gate-tunable SSH phase transition in superconducting resonators~\cite{splitthoff2024gate}, synthetic Rydberg atom arrays ~\cite{Kanungo_2022}, semiconductor quantum dots~\cite{kiczynski2022engineering},
superconducting metamaterials~\cite{jouanny2024band,mei2018robust,zheng2022engineering}, hybrid qubit-photon systems~\cite{kim2021quantum,vega2021qubit}. 

The paper is organized as follows:
in Sec.~\ref{sec:1D-chiral-models} we first review one dimensional chiral models for which we will analyze size-dependent phase diagrams given by the RSWN formula. Furthermore, we establish the criteria of bulk conductivity as a measure of the insulating/conducting state of the topological edge modes. In Sec~\ref{sec:rswn} we present the background and definition of the RSWN. In Sec.~\ref{sec:rswn-finite-example} we numerically illustrate an efficacy of bulk conductivity for the determination of the topological phases.
In Sec.~\ref{sec:analytic} we provide analytic insights into how edge states influence the phase diagram from a continuum model perspective. Finally, in Sec.~\ref{sec:disorder} we investigate the impact of disorder on the derived phase diagrams, examining how topological protection defined by RSWN is affected by various disorder. The conclusions are summarized in Sec.~\ref{sec:conclusions}.


\section{The Family of one dimensional Chiral symmetric models}
\label{sec:1D-chiral-models}
\subsection{Su–Schrieffer–Heeger (SSH) model}

Inspired by the modeling of polyacetylene, a long chain of carbon atoms with alternating single and double bonds, the Su–Schrieffer–Heeger (SSH) model \cite{su1979solitons, hasan2010colloquium, asboth2016short} has gained significant attention due to its profound connection between topology and physics.   

\begin{figure}
    \centering
    \includegraphics[width= 0.45\textwidth]{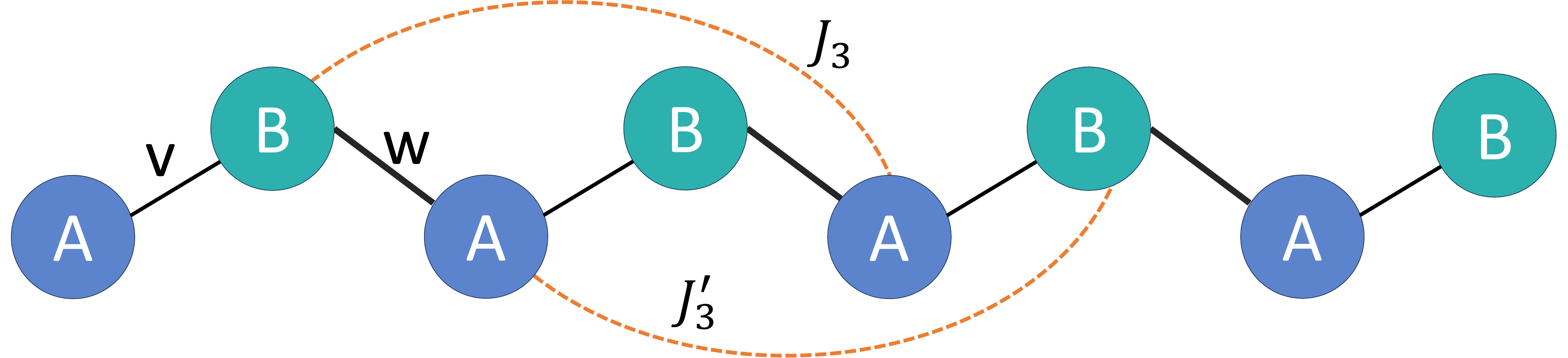}
    \caption{Lattice geometry of SSH model and extended SSH model. The SSH model contains only the nearest-neighbor hoppings, represented by $v$ and $w$ as solid lines. The extended SSH model contains the nearest-neighbor hopping and the third order hopping, represented by $J_3$ and $J_3^{'}$ in dashed orange lines.}
    \label{fig:Extend_SSH_schematic}
\end{figure}

\begin{figure}[h]
    \centering
    \includegraphics[width=0.475\textwidth]{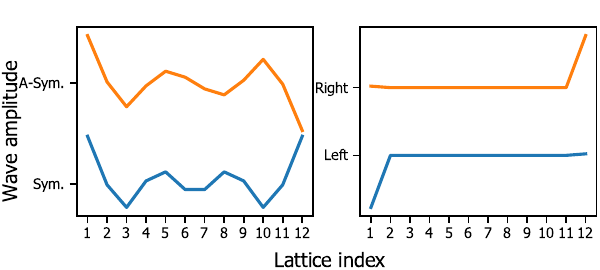}
    \caption{Example of hybridized SSH edge states (Left) with $v/w = 0.5, N=6$ and localized edge states (right) with $v/w =0.0005, N=6$. The x-axis label of odd (even) number represents type $A$ ($B$) sub-lattices. A-Sym.(Sym.) represents the anti-symmetric (symmetric) hybridization of two edges.}
    \label{fig:loc_hyb_states}
\end{figure}

The Hamiltonian of the SSH model is given by,
\begin{equation}\label{eq:ssh_eq}
    H=\sum_{n=1}^{N}(v \, c^{\dagger}_{n,A}c_{n,B}+ w\,c^{\dagger}_{n,B}c_{n+1,A}
    )+h.c.,
\end{equation}
where $c^{\dagger}_{n,\alpha}$ ($c_{n,\alpha}$) denotes the creation (annihilation) of a particle at lattice site ($n,\alpha$), with the unit cell index $n \in [1,N]$ and the sublattice index $\alpha \in {A,B}$. The ratio of intra-cell to inter-cell hopping, $v/w$, controls the topological phase transition. 
The SSH model hosts two topologically distinct phases: winding number is $\nu = 1$ when $v<w$, which is called the topologically non-trivial (topological phase), while for $v>w$, $\nu = 0$ which is called the topologically trivial (trivial phase).  The SSH model is classified as class AIII for spinful fermions or BDI for spinless bosons \cite{kane2014topological}. In the topologically non-trivial phase, two  mid-gap edge states arise, which are exponentially localized at the edge lattices, as shown in Fig.~\ref{fig:loc_hyb_states}.

The momentum-space Hamiltonian $H(k)$ of a two band model can be expressed in terms of Pauli matrices $\hat{\bm{\sigma}}$ as $H(k) = d_0(k)\hat{\sigma_0} +\bm{d}(k) \hat{\bm{\sigma}}$. The coefficients constitute the $\mathbf{d}$-vector, often referred to as the Bloch vector. The SSH Hamiltonian corresponds to $d_x(k) = v + w\text{cos}(k); d_y(k) = -w\text{sin}(k); d_z(k) = d_0(k)= 0$.

\subsection{Extended SSH model}

The chiral symmetry permits more types of hopping, provided the hoppings connect different sublattices (for example A-B and B-A). These types of hoppings are called odd hoppings. The most straight forward generalization of the SSH model is the extended SSH model with the third order hoppings (next next neighbor hopping). The Hamiltonian of the extended SSH model is given by,
\begin{equation}\label{eq:ext_ssh_eq}
\begin{split}
    H= \sum_{n=1}^{N}(
    &v \, c^{\dagger}_{n,A}c_{n,B}+ w\,c^{\dagger}_{n,B}c_{n+1,A} \\
    + &J_3 c_{n,B}^{\dagger} c_{n+2,A}
    + J_3^{'} c_{n,A}^{\dagger} c_{n+1,B}
    )+h.c.\\
\end{split}
\end{equation}
In addition to the SSH model, Eq.~\eqref{eq:ssh_eq}, $J_3$ and $J_3^{\prime}$ are the third order hoppings,  represented by orange dashed line in Fig ~\ref{fig:Extend_SSH_schematic}. When both first and third-neighbor hoppings are considered, the system can be configured to preserve time-reversal, particle-hole and chiral symmetry, thus it belongs to the BDI class, just as the standard SSH model. The corresponding topological phases are characterized by winding number $|\nu| = 0, 1, 2 $ in this model.

In models with long-range odd hoppings, the symmetries of the standard SSH model are maintained. However, the extended SSH models allow for larger values of the topological invariant. For a given hopping order $N$, the maximum possible winding number is $\nu_{\text{max}} = (N + 1)/2$. The winding number also indicates the maximum number of edge mode pairs that can exist in the system, reflecting the expanded topological behavior of models with higher-order hopping terms~\cite{perez2018ssh}.

Switching to the momentum space, the $d$-vector of the extended SSH model with third order hopping is be given by\cite{perez2018ssh},
\begin{equation}
\begin{split}
     d_x(k) & = v + (w + J_3') \, \text{cos}(k) + J_3 \, \text{cos}(2k),\\ 
     d_y(k) & = (w - J_3')  \, \text{sin}(k) + J_3 \, \text{sin}(2k), \ \ d_z(k) = 0.
\end{split}
\label{eq:exdent_dvec}
\end{equation}
We will calculate the k-space winding number later with this expression.

The extended SSH model can support up to two pairs of edge states, corresponding to the highest winding number $\nu = 2$. These edge states are characterized by their wave function amplitudes, which are localized at the left or right edges of the system and on either the type-$A$ or type-$B$ sublattices. The exact position of the maximum amplitude depends on the tuning parameters $v, w, J_3$, and $J_3^{\prime}$. Consequently, the edge modes can be predominantly localized at the first, third, or fifth site of the edge of the chain~\cite{perez2018ssh}.

\begin{figure}[b]
    \centering
    \includegraphics[width= 0.475\textwidth]{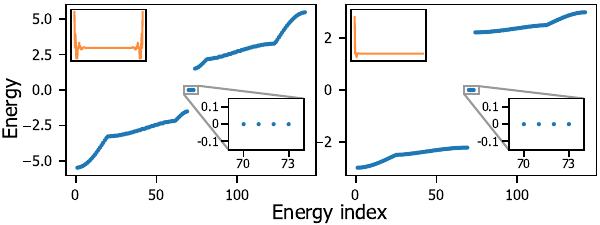}
    \caption{Energy spectrum of extended SSH models. The system size is $N=71$.
    Left: $v=0.5,w=1,j_3=3,j_{3p}=1,N=71$. Right: $v=0.25,w=0.25,j_3=2.5,j_{3p}=0.0,N=71$. Both panels are in the $\nu = 2$ phase. 
    The inset of each panel shows a zoom in of the zero energy region. Each insets shows four nearly degenerate mid-gap modes as a manifestation of $\nu=2$ phase. The corresponding mid-gap state wavefunctions are plotted in Fig.~\ref{fig:exdssh_edgestates} respectively.}
    \label{fig:essh_spectrum_two_peak_combi}
\end{figure}
 
\begin{figure}[h]
    \centering
    \includegraphics[width=0.5\textwidth]{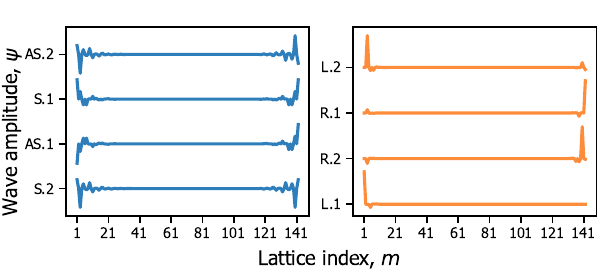}
    \caption{Examples of edge states in Extended SSH model. $N=71$.
    Left panel: hybridized edge states derived from $v=0.5,w=1,j_3=3,j_{3p}=1$.    
    Right panel: localized edge states derived from $v=w= 0.25,j_3=2.5,j_{3p}=0$.
    Lattice index $m$ represents sub-lattices. 
    In y-axis label, AS (S) represents anti-symmetric (symmetric) edge states, while
    L (R) represents the localized edge states on left (right) edges. Similarly, 1 (2) represents the location of the peak on first (second) edge lattice unit cells.
    E.g., AS.1 represents the first anti-symmetric mode (peak on the edge unit cell). 
    S.2 represents the second symmetric edge state (peak on the second unit cell from the edge). 
    }
    \label{fig:exdssh_edgestates}
\end{figure}

\subsection{Bulk conductivity of topological edge modes}

The successful application of topological materials relies on the bulk conductivity of topological edge modes in transport experiments~\cite{splitthoff2024gate}. This requires the two edge modes have non-negligible overlaps in the bulk so they can hybridize. However, in theory the topological edge states do not inherently guarantee conductivity, for example in Fig~\ref{fig:loc_hyb_states}, the right panel shows the localized edge states, with negligible overlap in the bulk - this type of edge modes does not have a conducting bulk and therefore the transport measurement will yield an insulator. The insulating edge states usually appear in larger systems or in systems deep in the topological phase. In our numerical analysis, we establish a criterion by putting a threshold for the averaged amplitude across the central two unit cells, defined as follows:

\begin{equation}                                         (|\Psi_{N/2,A}|+|\Psi_{N/2,B}|+|\Psi_{N/2+1,A}|+|\Psi_{N/2+1,b}|)/4 < 10^{-15}.
\label{eq:criteria} 
\end{equation}
Edge states that meet (or fail to meet) this criterion are classified as bulk insulators (or conductors), respectively. The threshold value is set to be $10^{-15}$ due to numerical precision of the python package 'NumPy' \cite{harris2020array}. The data type 'float64', default for most NumPy calculations, has about 15–16 decimal digits of precision. This is the level we consider to be indistinguishable numerically.

In the standard SSH model, we observe that edge states satisfying bulk conductivity condition, Eq.~\eqref{eq:criteria}, are consistently localized on a single edge, while edge states that do not satisfy Eq.~\eqref{eq:criteria} exhibit hybridization across both edges. Conversely, in the extended SSH model, we find that the localization and hybridization of pairs of edge states show no straightforward dependence on whether the criteria in Eq.~\eqref{eq:criteria} are met or violated.
Therefore, we expect that, rather than edge hybridization, the bulk conductivity condition Eq.~\eqref{eq:criteria} is closely related to the finite size topology. We will elaborate on this distinction etween models and validate our assumption in the following sections.


\section{Covariant representation of Winding number in real space}
\label{sec:rswn}

The covariant real space representation of winding number was first proposed in Ref.~\cite{mondragon2014topological}. The resulting formula has been corroborated in several subsequent studies ~\cite{song2014aiii,rakovszky2017detecting,prodan2016bulk}. In particular, this real space formulation can be useful for disordered systems that respect the underlying symmetry. The real space topological invariant defined in Eq.~\eqref{eq:rswn2} is a global quantity, and importantly, it retains its quantization even in the presence of strong disorder~\cite{mondragon2014topological, song2014aiii,prodan2016bulk}.

While the calculation of the RSWN is often presented in a complex form, we offer a streamlined and pedagogical approach to make this calculation more accessible. For a generic chiral symmetric system, we begin by recalling the anti-commutation relation between the Hamiltonian $H$ and the chiral operator, which satisfies $\Gamma$ $\{\Gamma, H\} = 0$. As a consequence,  the Bloch Hamiltonian $H(k)$ can be written in an off-diagonal form when expressed in the basis of the chiral symmetry operator, where the operator assumes the form of $\sigma_z$ in the case of SSH model)~\cite{schnyder2009lattice}.
$$
H(k) = \begin{pmatrix}
    0 & h(k)\\
    h(k)^{\dagger} & 0 \\
    \end{pmatrix},
$$
where $h(k) = d_x(k) - i d_y(k)$.

In parallel, the original formulation of the k-space winding number is given by~\cite{schnyder2009lattice}
\begin{equation}\label{eq:kspacewinding}
    \nu = \frac{1}{2\pi i } \int_{BZ} \text{Tr} \{ [h(k)^{-1} \partial_{k} h(k)] \} \in \mathbb{Z}.
\end{equation}

The expression in Eq.~\eqref{eq:kspacewinding} has a direct covariant form in real space~\cite{mondragon2014topological}. Starting from the real space SSH Hamiltonian $H$, the homotopically equivalent flatband version of $H$ is given by
\begin{equation}\label{eq:flatbandhamiltonian}
    Q = \frac{H}{|H|} = 
    \begin{pmatrix}
        0 & Q_0 \\
        Q_0^{\dagger} & 0 \\
    \end{pmatrix},
\end{equation}
where the off-diagonal matrix $Q_0$ will enter the real space winding number calculation. The covariant real-space form of the winding number can be obtained by applying the Bloch-Floquet transformation of Eq.~\eqref{eq:kspacewinding} written in terms of flatband Hamiltonian~\cite{mondragon2014topological}
\begin{equation}\label{eq:rswn}
    \nu = - \text{Tr}_{\text{volume}}\{ Q_0^{-1} [X, Q_0]\},
\end{equation}
where $X$ is the position operator and $\text{Tr}_{\text{volume}}$ denotes the trace per volume.  In the context of SSH model, position eigenstates are $|x,c \rangle$ are labeled by the unit cell index $x = 1, \ldots ,L$ and the sublattice degree of freedom $c = A, B$. The position operator $X$ therefore takes the form 
\begin{equation}
    X = \sum_{x\in \mathbb{Z}} \sum_{c=A,B} x |x,c\rangle \langle x,c|.
\end{equation}

The primary task is obtaining the off-diagonal term $Q_0$. Note that the chiral symmetry for lattice Hamiltonians is also referred to as sublattice symmetry. It can be defined by grouping all internal states into two sets, the sublattices $A$ and $B$. In order to obtain $Q_0$, we begin by defining the spectral projectors of chiral symmetry as follows:
\begin{equation}
\begin{split}
    P_A =&\sum_{x\in \mathbb{Z}} \sum_{c=A} | x,c\rangle \langle x,c|,\\ 
    P_B =&\sum_{x\in \mathbb{Z}} \sum_{c=B} | x,c\rangle \langle x,c| = \hat{I} - P_A.
\end{split}
\end{equation}
The chiral operator $\Gamma$ is given by 
\begin{equation}
    \Gamma = P_A- P_B.    
\end{equation}
Chiral symmetry anti-commutes with the Hamiltonian,meaning the Hamiltonian has no matrix elements between states on the same type of sublattice. $\Gamma$ acts on each unit cell separately and it satisfies the unitary condition $\Gamma^{\dagger} = \Gamma^{-1} = \Gamma$. An immediate consequence of chiral symmetry is that the eigenstates come in pairs $\{ |n\rangle, \Gamma|n\rangle \}$, with energies $\{-E_n,E_n\}$. To complement this framework, we introduce another set of projectors that partition the energy spectrum into upper and lower halves, given by,
\begin{equation}
\begin{split}
    P_{-} &= \sum_{E_n \leq 0} |n\rangle \langle n|, \\
    P_{+} &= \sum_{E_n \geq 0} |n\rangle \langle n| =\Gamma P_{-} \Gamma.
\end{split}
\end{equation}
Recall the definition of the flatband Hamiltonian $Q$ in Eq.~\eqref{eq:flatbandhamiltonian} and note that in the eigenstates basis it can be expressed as
\begin{equation}
    Q = \frac{H}{|H|} = \frac{\sum_{n} E_n|n\rangle \langle n|}{\sum_{n}E_n} = \sum_{n} |n\rangle \langle n|.
\end{equation}
Therefore, the flatband Hamiltonian in Eq.~\eqref{eq:flatbandhamiltonian} can be written as
\begin{equation}
     Q = P_{+} - P_{-}.
\end{equation}
As a chiral-symmetric operator, $Q$ satisfies the condition
\begin{equation}
    Q = P_A Q P_B + P_B Q P_A,
\end{equation}
the off-diagonal terms in Eq.~\eqref{eq:flatbandhamiltonian} can be written in terms of the above defined operators as 
\begin{equation}
    Q_{0} = P_A Q P_B, \quad (Q_{0})^{-1} = P_{B} Q P_A.
\end{equation}
Finally, one can express real space winding number Eq.~\eqref{eq:rswn} as 
\begin{equation}\label{eq:rswn2}
    \nu = - \frac{1}{L} \text{Tr} \left\{ P_{B} Q P_A \left[X, P_A Q P_B \right] \right\}.
\end{equation}

This real space representation reduces to the conventional k-space form in the presence of translational invariance, and it remains valid and quantized event in the presence of disorders~\cite{mondragon2014topological,song2014aiii}. 
In the subsequent discussion, we will refer to Eq.~\eqref{eq:rswn2} as the real-space winding number (RWSN). This formula is known to distinguish topological phases in chiral symmetric system like SSH model\cite{su1979solitons} and the Kitaev
chain\cite{kitaev2001unpaired}.

The RSWN ~\eqref{eq:rswn2} serves as a global real space topological invariant. There also exist well-defined local topological invariants. The topological winding marker\cite{meier2018observation}/ Chern maker\cite{bianco2011mapping,caio2019topological} are such invariants and they can locally distinguish topological phases in 1D/2D. The 1D winding marker has a very similar constitution to RSWN, given by \cite{meier2018observation}
 \begin{equation}\label{eq:marker}
     w(x) = \frac{-1}{V_{uc}} \sum_{\alpha}\bra{x,\alpha}Q_0^{-1} [X,Q_0]\ket{x,\alpha},
 \end{equation}
where $x$ are positions, $\alpha$ are sublattice indices and $V_{uc}$ is the unit cell volume.


\section{RSWN and bulk conductivity in finite size systems}
\label{sec:rswn-finite-example}

\subsection{RSWN in SSH model}

In many cases, the SSH Hamiltonian itself already reveals most information we need. However, this is not the case when one thinks about the bulk conductivity of the edge states. One needs to diagonalize the Hamiltonian and check the mig-dap modes in order to obtain that bulk conductivity information. We find interesting relationship between the finite size topology and the bulk conductivity, which we elaborate in the following.

\begin{figure}
    \centering
    \includegraphics[width=0.475\textwidth]{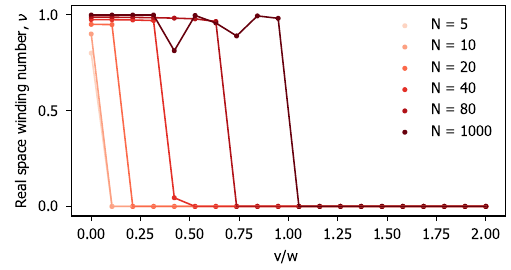}
    \caption{Real space winding number(RSWN) $\nu$ calculated for different system sizes $N \in \{5,1000\}$. Each dot represents the RSWN $\nu$ of an SSH Hamiltonian rendered at corresponding $v/w$ value on the $x$-axis. The Hamiltonian parameters are set to be  $w = 0.5$ and $v \in \{0, 1.0\}$.}
    \label{fig:rswn_vw_final}
\end{figure}

When calculating the RSWN $\nu $  from Eq.~\eqref{eq:rswn2} for finite-size SSH chains, the results reveal a deviation from the expected topological phase transitions. Figure ~\ref{fig:rswn_vw_final} shows the RSWN Eq.~\eqref{eq:rswn2} obtained from SSH chain of size $N\in \{5, 1000\}$. Interestingly, the RSWN transition occurs at a smaller $v/w$, rather than at $v/w = 1$. For $N = 5$, RSWN $\nu = 0$ in most cases where it is expected to be $\nu = 1$. However, for $N=1000$, the transition happens at $v/w = 1$, as predicted by the winding number formulated in k-space. This example illustrates that in finite size systems, the RSWN, as the counterpart of k-space winding number in real space, becomes unreliable in the topologically non-trivial regime. 

\begin{figure}
    \centering\includegraphics[width=0.475\textwidth]{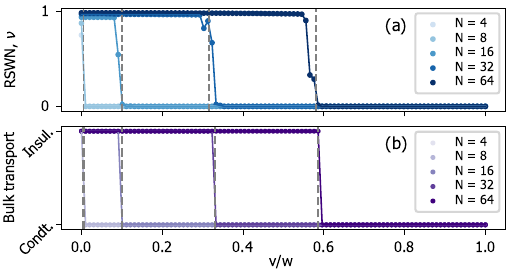}
    \caption{(a) Real space winding number(RSWN) $\nu$ calculated for small system sizes. Each dot represents the RSWN $\nu$ of an SSH Hamiltonian rendered at corresponding $v/w$ value on the $x$-axis with $v/w \in \{0, 1.0\}$. The topological transition is marked by a dashed vertical line in different system size $N$. (b) Bulk transport properties of the SSH topological edge states. $y$-axis shows the two states of the edge mode: Conduction (Condt.) and insulation (Insul.). The criterion of transition from conduction to insulation is the overlapping of wave function amplitude at the system bulk. Our definition in Eq.~\eqref{eq:criteria} is: the averaged probability amplitude of middle two unit cells $< 10^{-15}$. Python package `numpy` and `numpy linalg eig`are used. }
    \label{fig:SSH_condt_insul_combi}
\end{figure}

In order to investigate the anomalous values in RSWN, we analyze the bulk transport properties of two topological edge modes, defined in Eq.~\eqref{eq:criteria}. We first observe that the bulk conductivity of
the two edge modes from the same Hamiltonian remains identical, meaning they either conduct or insulate under the same parameter set. 
Additionally, we find that the transition in bulk transport aligns with the transition in RSWN, as shown in Fig.~\ref{fig:SSH_condt_insul_combi}, with both transitions exhibiting a clear dependence on system size. This phenomenon will be further analyzed in the following sections. 
Notably, several methods have been proposed to address the anomalous RSWN values. For example, Song, F., et al.~\cite{song2019non} considered taking the trace only over the bulk. Similarly, winding marker Eq.~\eqref{eq:marker} incorporates even more localized information.

\subsection{RSWN in the extended SSH model}

\begin{figure}[h]
    \centering
    \includegraphics[width= 0.475\textwidth]{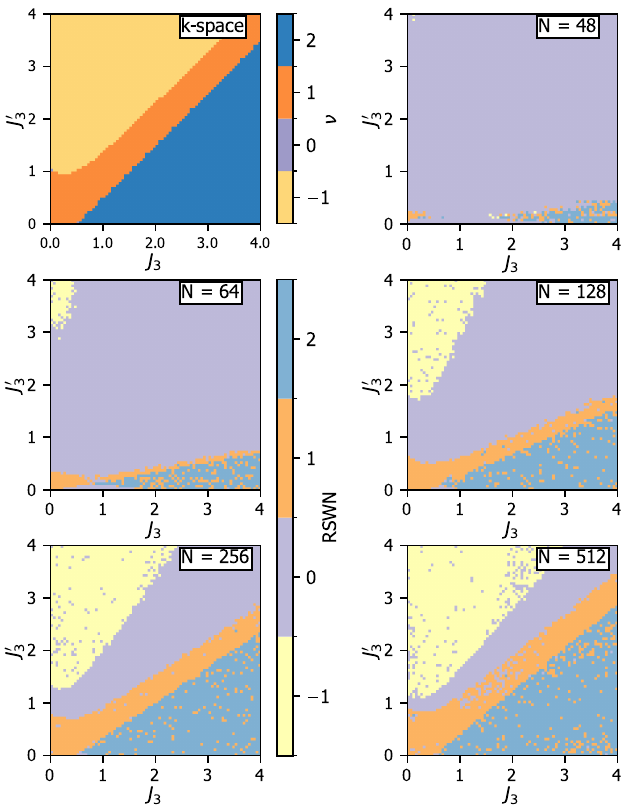}
    \caption{Winding number calculated from the Extended SSH model in finite size systems and the theoretical k-space limit. The parameters $v = 1/2, w =1$ are set to be fixed  values. $J_3$ and $J_3\prime $ are varied in the range of $[0, 4]$. }
    \label{fig:RSWN_6figs}
\end{figure}

In the previous section, we demonstrated that the RSWN becomes unreliable in the topologically non-trivial regime for the SSH model. A natural question arises: how does this formula perform in systems with multiple topological phases, such as those with higher winding numbers?
In order to answer this question, we apply the same RSWN formula Eq.~\eqref{eq:rswn2} to the extended SSH model, which includes third order hopping, allowing the winding number to reach $|\nu| = 2$. The corresponding winding number is calculated and shown in Fig.~\ref{fig:RSWN_6figs}. We set the $v,w$ parameters to be fixed at $v = 1/2, w =1$ and $J_3$ and $J_3\prime $ are varied in the range of $[ 0,4]$. For small system size, such as $N=48$, the RSWN prediction can be completely inaccurate. It still restores the k-space prediction as the large system sizes grows to $N=512$. We notice that there is a $\nu=0$ region even in the $N=512$ case that is absent in k-space plot. This indicates that higher ordered topological phases are more susceptible to finite size effect.

\begin{figure}[h]
    \centering
    \includegraphics[width=0.475\textwidth]{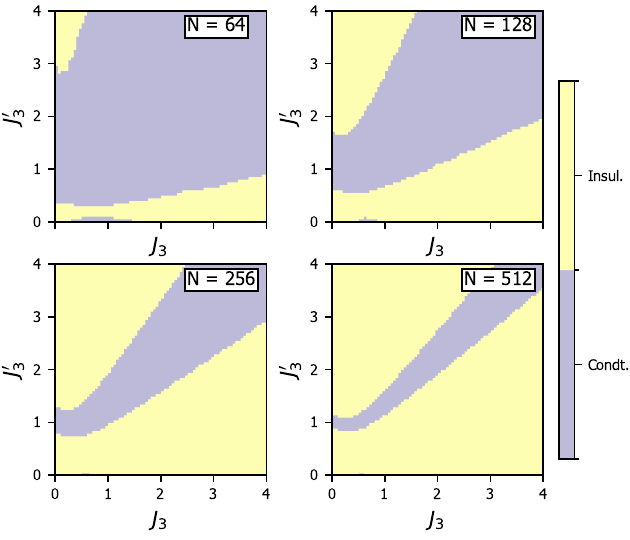}
    \caption{Mid-gap edge states bulk conductivity phase diagram in the extended SSH model of finite size systems. The parameters $v = 1/2, w = 1$ are set to be fixed  values. $J_3$ and $J_3\prime $ are varied in the range of $[0, 4]$. Label \textbf{Insul.} represents the edge state with no overlap in the bulk, while label \textbf{Condt.} represents the case where there exists non-negligible overlap between two edges in the topological edge states.}
    \label{fig:Hyb_trans_4figs}
\end{figure}

Again, we examine the bulk transport properties of sets of mid gap topological edge states in the extended SSH mode, the result is shown in Fig.~\ref{fig:RSWN_6figs} and Fig.~\ref{fig:Hyb_trans_4figs}. The key message from these two figures is that, although the RSWN is affected by finite system size, it still reflects the bulk transport behavior. In Fig.~\ref{fig:Hyb_trans_4figs}, the purple region characterizes the conducting bulk, which perfectly overlaps with the RSWN $\nu = 0$ region in  Fig.~\ref{fig:RSWN_6figs}. Likewise, the yellow region in Fig.~\ref{fig:Hyb_trans_4figs}, characterizing the bulk insulating phase, fully overlaps with the non-zero RSWN region in  Fig.~\ref{fig:RSWN_6figs}.

In conclusion, our numerical study strongly indicates that the bulk conductance serves as an experimentally accessible indicator of the RSWN. In the following section, we will support our reasoning by an analytical analysis.


\section{Analytical effective description of RSWN phase diagrams}
\label{sec:analytic}
In this Section we present analytic explanation of phase diagrams obtained above. The calculations are based on perturbative estimation of overlap between edge states.

We start from SSH model, where all calculations could be performed exactly. At phase transition point $v=w$ the gap closing for translation-invariant model happens at the points $k=\pm\pi$ of the first Brillouin zone. Decomposing the tight-binding Hamiltonian up to linear order in $q$ for  $k=\pi+q$, we obtain:
\begin{align}
    H_{lin,\pi}(k)=\left(\begin{array}{cc}
0 & v-(1+i q) w \\
v-(1-i q) w & 0
\end{array}\right).
\end{align}
This Hamiltonian describes effective low-energy model, which predicts linear dispersion of quasi-particles at phase transitions $v=w$. Away from phase transition, the dispersion $\pm \sqrt{(v-w)^2+w^2q^2}$ has a gap of size $2\Delta$ with $\Delta=|v-w|$. To find the eigenstates in real space system, we transform to coordinate space representation by substituting $q\to -i\partial_x$. Then, the Schroedinger equation $H\Psi=E\Psi$ becomes a system of two coupled differential equations for $\Psi=(a(x), b(x))$:
\begin{align}
\begin{cases}
        (v-w) b(x)+w \dot{b}(x)=E a(x),& \\
        (v-w) a(x)-w \dot{a}(x)=E b(x).& 
    \end{cases}
\end{align}
The general solution of this system of differential equations for $E\neq 0$ reads
\begin{align}
    a(x)&=C_1 \sin \left[\sqrt{-(v / w-1)^2+E^2/w^2} x\right]\nonumber\\
    &+C_2 \cos \left[\sqrt{-(v / w-1)^2+E^2/w^2} x\right],\nonumber\\ 
    b(x)&=\frac{(v-w) a(x)-w \dot{a}(x)}{E}.
\end{align}
The boundary conditions in the finite system of length $L$ (even number, $L=2N$ with $N$ being the number of unit cells) are $a(0)=0$, and $b(L)=0$. The exactly zero energy solutions of this system of equation do not satisfy these boundary conditions, since they have the form $a(x)=C_a e^{(v / w-1) x}, \,\, b(x)=C_b e^{-(v / w-1) x}$.
Substituting general solution into boundary conditions, we get a spectral equation: 
\begin{align}
\text{tan} \left[\sqrt{-\left(\frac{v}{w} -1\right)^2+\frac{E^2}{w^2}} L\right]=\frac{\sqrt{-(v -w)^2+E^2}}{w-v}.
\end{align}
To find the solution for edge states within the gap range $|E|<\Delta$ in the topological phase with $w>v$, we perform a series expansion of both sides up to leading order in $|E|/\Delta$. Approximating the square root 
\begin{align}
    \sqrt{-(v -w)^2+E^2}\approx i|v -w|\left(1-E^2 / 2\Delta^2\right),
\end{align}
we find the solution for the energy of edge states:
\begin{align}
    \label{eq:overlap-value-approx}
    E_0^{\pm}&\approx\pm \sqrt{\frac{2   \Delta^2 \left(e^{-2L\frac{\Delta}{w}}+1\right)}{\cosh \left(2 L \frac{\Delta}{w}\right)-2 L
   \frac{\Delta}{w} +1}}\sim 
\pm  \sqrt{2}\Delta e^{-L\frac{\Delta}{w}}.
\end{align}
The solutions that correspond to these energies have approximate form $\Psi_{\pm E_0}(x)=(a(x),b(x))$ with
\begin{align}
    &a_{\pm E_0}(x)=C_1\sinh[\sqrt{\Delta^2-E_0^2} x/w],\\ &b_{\pm E_0}(x)=\pm \frac{C_1}{|E_{0}|}\left(\sqrt{\Delta^2-E_0^2}\cosh[\sqrt{\Delta^2-E_0^2} x/w]+\right.\nonumber\\
    &\left.\Delta\sinh[\sqrt{\Delta^2-E^2} x/w]\right).
\end{align}
These solutions are symmetric and antisymmetric and are localized near both edges. Thus, in the continuum theory there are no states localized only on single edge. To estimate the contribution of such states into RSWN formula, we substitute only these states into Eq.\eqref{eq:rswn2}. Due to particle-hole symmetry, the states' vectors are $|\pm E_0\rangle=(a(x),\pm b(x))$ and the $Q$-operators read $Q=|+E_0\rangle\langle+E_0|-|-E_0\rangle\langle-E_0|$. Applying the projectors on sublattices, we obtain expression for trace
\begin{align}
    &\text{Tr} P_B Q P_A\left[X, P_A Q P_B\right] = \text{Tr}(\left|0,b\right\rangle\left\langle a,0\right|-\left|0,-b\right\rangle\left\langle a,0\right|)\times\nonumber\\
    &\left(\hat{X}  (\left|a,0\right\rangle\left\langle 0,b\right|-\left|a,0\right\rangle\left\langle 0,-b\right|)  -\right.\nonumber\\
    &-\left.(\left|a,0\right\rangle\left\langle 0,b\right|-\left|a,0\right\rangle\left\langle 0,-b\right|) \hat{X}\right)\nonumber\\
    &=\text{tr}(X_A)-\text{tr}(X_B),
\end{align}
where in the last line we evaluated all matrix elements between different bra and ket states, and introduced the notation for coordinate operator matrix elements:
\begin{align}
    &X_{A}=\langle a,0|\hat{X}|a,0\rangle=\operatorname{diag}\left(0, \ldots L a^2(L)\right)\\
    &X_{B}=\langle 0,b|\hat{X}|0,b\rangle= \operatorname{diag}\left(b^2(0), \ldots, 0\right).
\end{align}
The trace of the diagonal matrix in the first line creates a contribution $\sim -L/L$ that cancels all other bulk state contributions in the winding number. This demonstrates the general principle of how pairs of edge states with symmetric/antisymmetric shape on both sides of system reduce RSWN value by $1$. It is important to note that for localized edge states on one side, projection operation $P_A Q P_B$ directly gives zero, and thus RSWN value is not reduced. We note that since the values on the diagonal of matrices $X_{A,B}$ depend on exact distribution of $|a(x)|^2$ over coordinate space, the procedure of restoring back the value of winding number by removing few edge sites is not well defined and would depend on exact ratio of hopping parameters (as $v/w$ in SSH model)~\cite{song2014aiii}.

Thus, several elements at the start or end of the diagonal cancel out all other contributions of winding markers, resulting in zero value of winding number. To explain why the RSWN formula can be still used, the special attention should be paid to the meaning of topological protection in the real system.

The edge states are topologically protected if the crosstalk with the bulk states and between states is negligibly small comparing to the corresponding level broadening or coherence time.
The level broadening could happen when the temperature changes or due to the coupling to the environment.
In other words, if the energy $E_0\ll\Gamma$, with $\Gamma$ being a level broadening, both basis selections $\Psi_{\pm E_0}$ and $1/2(\Psi_{+E_0}\pm \Psi_{-E_0})$ are allowed. The second selection results in a value, predicted by RSWN formula, being close to k-space winding number up to $1/L$-corrections.
In numerical calculations with precision regulated by Eq.~\ref{eq:criteria}, the level broadening could be replaced by the accuracy of numerical solver. Together all these effects force the region of topologically trivial phase to spread into topologically nontrivial phase near the phase transition, effectively changing phase transition point in parameter space.

Having explained the general principle, we now extend the discussion to extended SSH model. Each phase transition line corresponds to gap closing at particular $k$-point, and these points are different for each transition. According to bulk-boundary correspondence, the change of winding number in translation-invariant model corresponds to the change of number of pairs of edge states in finite size systems.
Near the phase transition lines, the energy of edge states that disappear is determined similarly to Eq.~\eqref{eq:overlap-value-approx}. Thus, each pair of edge states that disappear lowers RSWN by 1.
This qualitative description covers all examples shown in Fig.\ref{fig:RSWN_6figs}. Note that the phases are still ordered by increasing winding number by the same steps as in k-space. That is because the gaps near single phase transition line could be ordered as $\Delta_0<\Delta_1\leq\Delta_2\dots$ where $\Delta_0=0$ exactly at phase separation line. 
While one pair of edge states that corresponds to $\Delta_0$ contributes to decreasing RSWN, the remaining ones are not. That is because their overlap in the bulk, governed mainly by exponential $e^{-\Delta L / w}$ as in Eq.\eqref{eq:overlap-value-approx}, is much lower than the level broadening and does not lead to hybridization. 

Before proceeding to analysis of the role of disorder, we note that the calculation of RSWN with the level broadening estimated for particular system might lead to the understanding of the quality of topological protection. In particular, while some pairs of edge states might have a crosstalk over bulk, the rest are still localized and would not loose the coherence.  This could be applied to extended SSH model with phase diagrams presented in Figs.\ref{fig:RSWN_6figs},\ref{fig:Hyb_trans_4figs}, where one should expect topological protection only for very small range of parameters in small system $N=48$, and different numbers of topologically protected states in longer systems. 

\section{Disorder}
\label{sec:disorder}
\subsection{Disorder in hopping parameters}

Topological materials are well known for their robustness against disorder, a property rooted in the topological protection derived from symmetries in momentum space. Since disorder breaks translational invariance—a key condition for defining topological invariants in momentum space—this robustness cannot be directly demonstrated through standard momentum-space calculations. However, using the RSWN, we can capture this robustness in finite-sized disordered systems.

In the SSH family, there are two types of disorders: hopping disorders, which preserves the chiral symmetry; and the disorders in the chemical potentials, which break the chiral symmetry \cite{asboth2016short}. We begin by discussing hopping disorders. Previous studies have shown that the RSWN calculated for hopping disordered SSH model retains the discrete values expected by it's k-space counterpart \cite{splitthoff2024gate, song2014aiii, prodan2016bulk}.

\begin{figure}[h]
    \centering
    \includegraphics[width=0.475\textwidth]{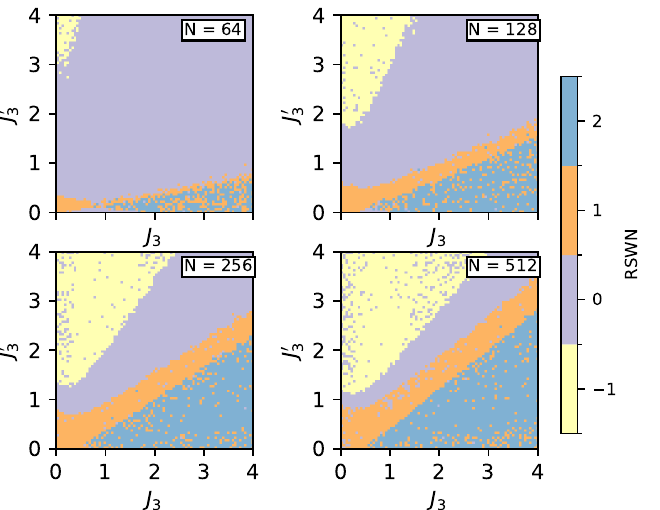}
    \caption{Real space winding number $\nu$ of extended SSH model with $5\%$ uncertainty in the hopping parameters $v, w, j_3, j_{3p}$.
    Similar to previous simulations, the parameters $v = 1/2, w =1$ are set to be fixed values. $J_3$ and $J_3\prime $ are varied in the range of $[0, 4]$.
    }
    \label{fig:esshdisorder}
\end{figure}

To further reinforce the confidence in topological physics, we investigate the behavior of RSWN on extended SSH under hopping disorders. Fig.~\ref{fig:esshdisorder} shows the phase diagrams of the extended SSH model with $5\%$ hopping disorders applied to the hopping parameters $v, w, j_3, j_{3p}$, i.e., the parameters fluctuate in the $\pm 2.5 \%$ region of their desired values. The RSWN remains  stable even with this $5\%$ uncertainty, as shown in Fig.~\ref{fig:RSWN_6figs}. In numerics, we set the designated values of hopping parameters to be: $v = 1/2, w =1$, while $J_3$ and $J_3\prime $ are varied in the range of $[ 0,4]$. The obtained phase diagrams truthfully resembles the original phase diagram in Fig.~\ref{fig:RSWN_6figs}.

\subsection{Disorder in chemical potential $\mu_i$}

A fundamental question in implementing topological phases of matter in real-world experiments is: \textit{What happens when perfect symmetry is slightly broken?} In the SSH model, the chiral symmetry requires equal local chemical potentials $\mu_i$, but achieving this exact condition in reality—whether in engineered meta-materials or natural condensed matter systems—is virtually impossible due to inevitable fluctuations. This raises the question: \textit{Does chiral symmetry ever truly exist in physical systems?} This is where the RSWN formula can give insight. 

\begin{figure}
    \centering
    \includegraphics[width=0.47\textwidth]{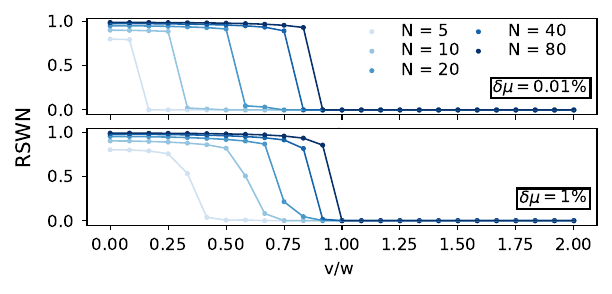}
    \caption{RSWN of SSH model vs $v/w$ when the chemical potentials $\mu_i$ are disturbed. Upper panel: the fluctuation in chemical potential is $\delta \mu = 0.01\%$. Lower panel: the fluctuation in chemical potential is $\delta \mu = 1\%$. The parameters are $v \in [0, 1], w = 0.5$.}
    \label{fig:rswn_chem_disorder_combined}
\end{figure}

We find that chemical potential disorder $\delta \mu$ can push the RSWN transition in finite size systems to the theoretical limit. For example, in Fig.~\ref{fig:rswn_vw_final}, the transition of $N=80$ occurs around $\sim v/w \sim 0.75$ in disorder free systems. However, when chemical potential disorder $\delta \mu$ is introduced. the transition of $N=80$ shifts closer to $v/w \sim 1$ as shown in Fig.~\ref{fig:rswn_chem_disorder_combined}.

\begin{figure}
    \centering
    \includegraphics[width=0.47\textwidth]{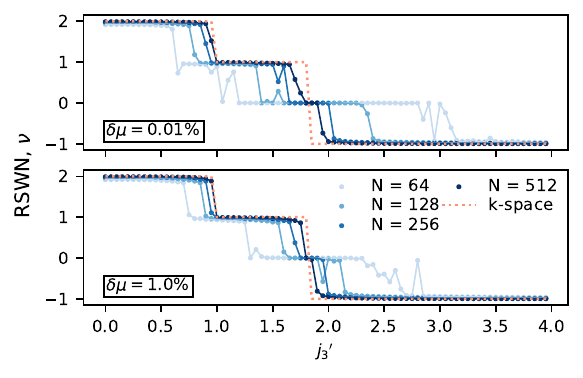}
    \caption{RSWN of Extended SSH model vs $j_3\prime$ when the chemical potentials $\mu_i$ are disturbed. Upper panel: the fluctuation in chemical potential is $\delta \mu = 0.01\%$. Lower panel: $\delta \mu = 1\%$. The theoretical phase transition predicted by k-space winding number is plotted in the orange dotted line. The parameters used are $v = 0.5, w = 1, j3 = 1.5$.}
    \label{fig:ESSH_rswn_chem_disorder_combined}
\end{figure}

Similarly, in the presence of chemical potential disorder $\delta \mu$, the phase transition in finite sized extended SSH model shifts closer to ideal values, as shown in Fig.~\ref{fig:ESSH_rswn_chem_disorder_combined}. Additionally, the anomalous $\nu = 0$ region (the plateau in Fig.~\ref{fig:ESSH_rswn_chem_disorder_combined}) is  narrowed. Thus, we conclude that the presence of chemical potential disorder $\delta \mu$ enhanced the observation possibility of topological phase transition in experimental implementations.

\section{Conclusion and discussion}
\label{sec:conclusions}

In this work, we addressed the critical limitations in the existing real space winding number Eq.~\eqref{eq:rswn2} based on one dimensional chiral symmetric models of different winding number thresholds. We demonstrated that while the RSWN deviates from expected values in small systems, it ultimately converges to theoretical predictions in larger systems.
We developed bulk conductivity criteria to reliably characterize the topology in finite sized and disordered topological systems. We showed that the combination of RSWN and the bulk conductivity serves as a good measure of topological protection. Our proposal compensates for the inaccurate prediction of topological phase transition from the RSWN alone.

We supported our conclusions with both numerical and analytical studies of the effect of edge states on the phase transition. The real space invariant together with edge states transport properties provides a wholesome overview of the topological properties of the system. Furthermore, we investigated the impact of hopping disorder and chemical potential disorder on RSWN predicted topological phase transitions. We numerically show that the RSWN is robust against symmetry respecting disorders. Additionally, we found the chemical potential disorder, which breaks the chiral symmetry, can shift the transition points closer back to the ideal values, thus supporting experimental implementations of topological materials.

In the near term, our results are immediately relevant to designing and optimizing topological materials for experimental setups that rely on robust edge states, such as quantum dots, superconducting resonators, and hybrid qubit-photon systems~\cite{Kanungo_2022,splitthoff2024gate,jouanny2024band,culcer2020transport,zheng2022engineering,bandres2018topological, kim2021quantum,freedman2003topological,dennis2002topological,stern2013topological,ten2024two}. It is worthy to note that the RSWN is usually hard to be measured directly from experiments, but can be learned with machine learning techniques \cite{caio2019machine}. Our proposed criteria involving bulk conductivity will help experimentalists verify and refine the topology of their systems even in disordered configurations. The potential experimental verification of, for example, the lifetime of topological modes could reveal the real life topology discussed in this work. Our result also opens the door to controlled disorder engineering as a tool to fine-tune topological phase transitions. This approach could significantly impact fields like quantum computing and topological quantum states, where scalable and resilient topological properties are crucial.

The future works can develop in the following two direction. First one can study the similar effect in higher dimensional topological materials, for instance two dimensional topological insulators~\cite{ten2024two,qi2009time}, and for different types of topological invariants~\cite{caio2019topological,bianco2011mapping}. Symmetries of different kinds should also be considered. In our case, the chiral symmetry is broken while the chemical potential is added, while in reality it does not completely destroy the topological phase transition. It is highly intriguing if the similar situation will happen when one moves to the topological crystalline insulator where the crystalline symmetry can be easily broken when one introduced local perturbations. We do believe the our transport-integrated criteria is vita when characterizing the finite sized systems. We think it is interesting to investigate new benchmarking method of real life topology.

Overall, our results suggest that the bulk conductivity is an indispensable compensation to the anomalous RSWN in finite sized systems with disorders. With the support from local bulk conductivity information, RSWN shows a clear robustness towards perturbations and suggests potential benefits of previously harmful disorders that disrespects the symmetry. Our work contributes to create a better designing principle of topological materials and a deeper understanding of topological physics. Our approach also aligns with the practical requirements of quantum information and quantum transport applications, where topological protection of edge states is essential.

\section{Data and Code Availability}

The code for generation and analysis of all phase diagrams and corresponding data files for extended SSH model can be found at the following repositories \cite{Supplement-code}.

\section{Author Contributions}
GJ conceived the project with input from DOO, LJS and EG. GJ performed the simulations and created the plots with input from DOO, LJS and EG. DOO performed the analytical calculations with input from GJ and EG. GJ, DOO, LJS and EG analyzed and interpreted the results and co-wrote the manuscript. EG supervised the project.

\section{Acknowledgments}

We are thankful for fruitful discussions with Ana Silva, Stan Bergkamp and Sibren van der Meer. 
This work is part of the project Engineered Topological Quantum Networks (Project No.VI.Veni.212.278) of the research program NWO Talent Programme Veni Science domain 2021 which is financed by the Dutch Research Council (NWO). This project was supported by the Kavli Foundation. This research was co-funded by the Dutch Research Council (NWO) and by the Top consortia for Knowledge and Innovation (TKI) from the Dutch Ministry of Economic Affairs. GJ acknowledges the research program “Materials for the Quantum Age” (QuMat) for financial support.

\nocite{*}
\bibliography{ref}
\end{document}